\documentclass{WileyMSP-template}

\usepackage{graphicx}%
\usepackage{multirow}%
\usepackage{amsmath,amssymb,amsfonts}%
\usepackage{amsthm}%
\usepackage{mathrsfs}%
\usepackage[title]{appendix}%
\usepackage{xcolor}%
\usepackage{textcomp}%
\usepackage{manyfoot}%
\usepackage{booktabs}%
\usepackage{algorithm}%
\usepackage{algorithmicx}%
\usepackage{algpseudocode}%
\usepackage{listings}%
\usepackage{soul}
\usepackage{float}
\usepackage{placeins}
\usepackage{flafter}
\usepackage{amsthm}%

\begin{document}

\title{XHEMTs on Ultrawide Bandgap Single-Crystal AlN Substrates}

\maketitle

% Author: Please give full first and last names for authors and include * after the name of all corresponding authors

\author{Eungkyun Kim*}
\author{Yu-Hsin Chen}
\author{Naomi Pieczulewski}
\author{Jimy Encomendero}
\author{David Anthony Muller}\\
\author{Debdeep Jena*}
\author{Huili Grace Xing*}

% Dedication

\dedication{}

% Affiliations: Please provide adacemic titles (Prof. or Dr.) for all authors where applicable, and include an institutional email address for all corresponding authors
\begin{affiliations}
Eungkyun Kim\\
Department of Electrical and Computer Engineering, Cornell University, Ithaca, NY 14853, USA.\\
ek543@cornell.edu

Yu-Hsin Chen, Naomi Pieczulewski\\
Department of Materials Science and Engineering, Cornell University, Ithaca, NY 14853, USA.

Dr. Jimy Encomendero\\
Department of Electrical and Computer Engineering, Cornell University, Ithaca, NY 14853, USA.

Professor David Anthony Muller\\
School of Applied and Engineering Physics, Cornell University, Ithaca, NY 14853, USA.

Professor Debdeep Jena, Professor Huili Grace Xing\\
Department of Electrical and Computer Engineering, Cornell University, Ithaca, NY 14853, USA.\\
Department of Materials Science and Engineering, Cornell University, Ithaca, NY 14853, USA.\\
Kavli Institute at Cornell for Nanoscale Science, Cornell University, Ithaca, NY 14853, USA.\\
djena@cornell.edu, grace.xing@cornell.edu

\end{affiliations}

% Keywords: Please provide a minimum of three and a maximum of seven keywords, separated by commas


% Abstract should be written in the present tense and impersonal style (i.e., avoid we), and be at most 200 words long
\begin{abstract}

AlN has the largest bandgap in the wurtzite III-nitride semiconductor family, making it an ideal barrier for a thin GaN channel to achieve strong carrier confinement in field-effect transistors, analogous to silicon-on-insulator technology. Unlike SiO$_2$/Si/SiO$_2$, AlN/GaN/AlN can be grown fully epitaxially, enabling high carrier mobilities suitable for high-frequency applications. However, developing these heterostructures and related devices has been hindered by challenges in strain management, polarization effects, defect control and charge trapping. Here, the AlN single-crystal high electron mobility transistor (XHEMT) is introduced, a new nitride transistor technology designed to address these issues. The XHEMT structure features a pseudomorphic GaN channel sandwiched between AlN layers, grown on single-crystal AlN substrates. First-generation XHEMTs demonstrate RF performance on par with the state-of-the-art GaN HEMTs, achieving 5.92 W/mm output power and 65\% peak power-added efficiency at 10 GHz under 17 V drain bias. These devices overcome several limitations present in conventional GaN HEMTs, which are grown on lattice-mismatched foreign substrates that introduce undesirable dislocations and exacerbated thermal resistance. With the recent availability of 100-mm AlN substrates and AlN’s high thermal conductivity (340 W/m$\cdot$K), XHEMTs show strong potential for next-generation RF electronics.

\end{abstract}

% Text: Please use section headings and subheadings as specified below. For communications, all section headings apart from Experimental Section should be removed
% Please make the first reference to a display item bold: \textbf{Figure 1}
% Do not abbreviate Figure, Equation, etc.; display items are always singular, i.e., Figure 1 and 2.
% Equations are always singular, i.e., Equation 1 and 2, and should be inserted using the {equation} environment, not as graphics
% Please do not use footnotes in the text, additional information can be added to the Reference list.

\section{Introduction}
\begin{figure}[h!]
\centering
\includegraphics[width=\textwidth]{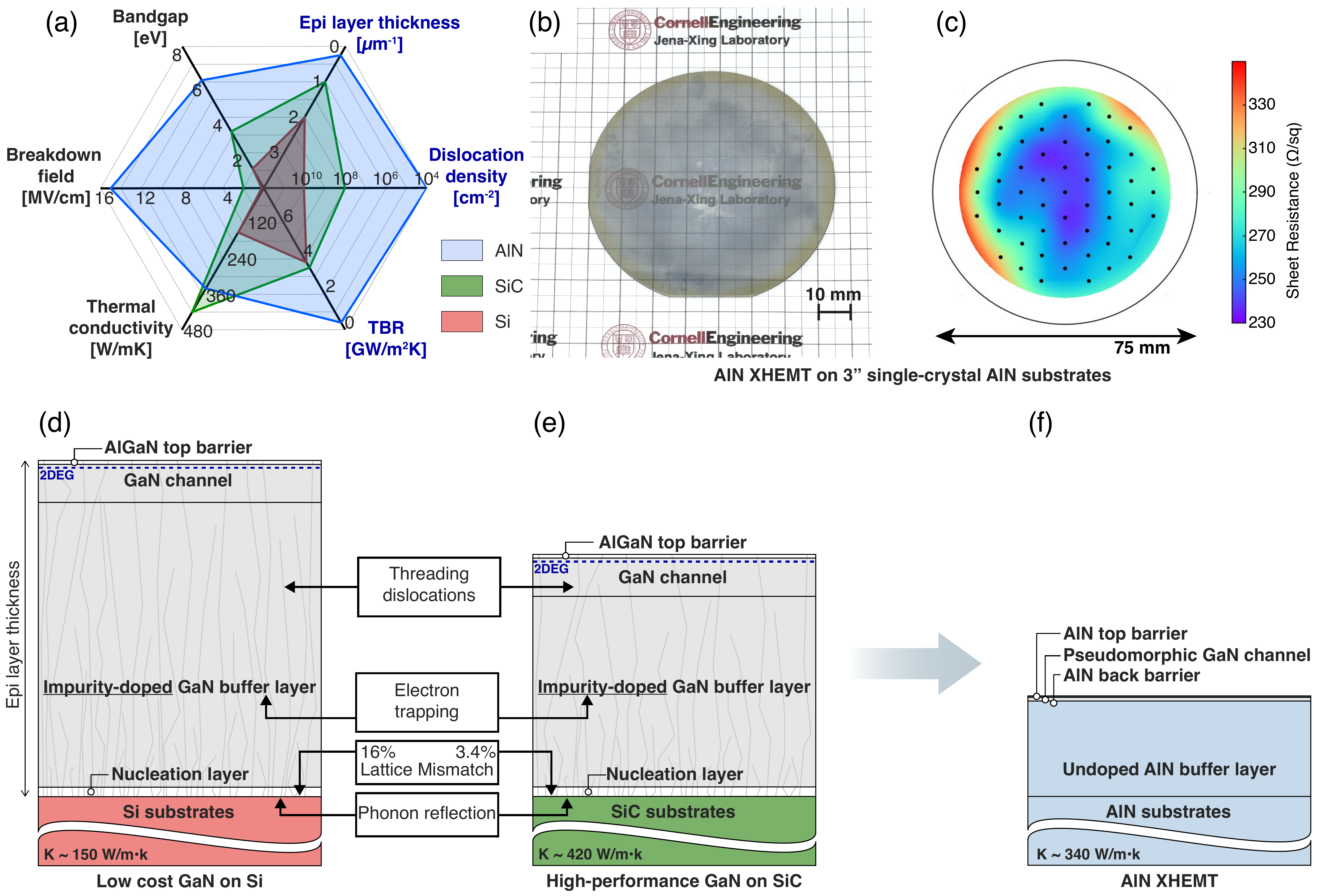}
\caption{A tale of three substrates for nitride HEMTs. (a) Spider plot comparing the key material properties of substrates and device epitaxy used in GaN HEMTs. The AlN/GaN/AlN XHEMT platform promises to take advantage of AlN's unique combination of an ultrawide bandgap and high thermal conductivity. More importantly, pseudomorphic epitaxy of ultra-thin GaN channel on single crystal AlN leads to very low dislocation densities (determined by the substrate, about $10^4$ cm$^{-2}$) and nearly zero thermal boundary resistance. (b) Optical image (c) and sheet resistance map of the AlN/GaN/AlN XHEMT grown on 3-inch single-crystal AlN substrates. Representative layer structures of commercially available GaN HEMTs on (d) Si substrates for cost-effectiveness and (e) SiC substrates for high-performance. Both are affected by electron trapping in the buffer layer, high dislocation densities (about $10^9$ cm$^{-2}$), and a high thermal boundary resistance (TBR) arising from heteroepitaxy, which also necessitates much thicker epitaxial buffer layers. These challenges are more pronounced in Si substrates due to their lower thermal conductivity, larger lattice and phonon mismatch with GaN. (f) AlN XHEMT structure addresses these issues by offering zero lattice and phonon mismatch at the growth interface on the substrate, ultra-thin channel with large electron confinement, and high thermal conductivity and high electrical resistivity of AlN.} 
\label{fig1}
\end{figure}

Often the development of new crystals creates entirely new fields, and simultaneously rejuvenates the existing state of the art. High quality single-crystal wafers of the ultrawide bandgap semiconductor aluminum nitride (AlN) up to 100 mm diameter in size have recently become possible \cite{CrystalIS100mm_Pssb, 2-inch_AlN}. Such substrates are unleashing a new era of UV photonics by enabling the first ever electrically injected continuous wave deep-UV semiconductor diode lasers \cite{DUV_Laser}. AlN has a direct energy bandgap $>$6 eV, an ultrahigh natural electrical resistivity of $\rho_{\rm 300K} \approx 10^{14}$ $\Omega \cdot {\rm cm}$, and a high thermal conductivity of $\kappa_{300 {\rm K}}\approx 340$ W/m$\cdot$K [see Figure \ref{fig1} (a)] \cite{Rounds_AlN_K}.  This combination is a `dream list' of desirable properties of ultrawide bandgap semiconductor substrates for next generation high speed microwave transistors for electronics, provided high electrical conductivity channels can be created and modulated with a gate voltage efficiently. 

To add to the above physical properties, the AlN crystal has a wurtzite lattice structure which exhibits broken inversion symmetry along the $c-$axis.  This broken symmetry produces large spontaneous and piezoeletric polarization, a physical property that is absent in other ultrawide bandgap semiconductors such as diamond and $\beta-$Ga$_2$O$_3$.  The change in polarization across junctions has enabled the formation of high mobility two-dimensional electron gases (2DEGs) \cite{Cindy_QW, Cindy_XHEMT, Qi_FirstAlN} and two-dimensional hole gases (2DHGs) \cite{ZZ_2DHG} at undoped quantum well (QW) heterojunctions of AlN and GaN grown on single-crystal AlN substrates with nearly {\em million times} lower dislocation densities than conventional templates such as AlN on silicon or SiC. 

Can the newly available large-area bulk AlN substrate platform rejuvenate nitride electronics?  In this work we show how a new device structure called the XHEMT [for single-crystal (Xtal) high electron mobility transistor] exploits the unique properties of the high quality single-crystal AlN bulk substrate to provide a path to overcome several hurdles faced in today's nitride transistors. We also explicitly point out the challenges that need to be overcome for the XHEMT in the near future.  By overcoming some hurdles, we demonstrate that XHEMTs deliver high currents, output powers of $\sim$ 6 W/mm, and 65\% power added efficiency in the X-Band, establishing it as a new nitride transistor technology.  The new XHEMT architecture enables a number of new possibilities that are not attainable in current AlGaN/GaN HEMTs, and thus offers an exciting future for nitride electronics.

Figure \ref{fig1}(b) shows an example 75 mm single-crystal Al-polar AlN bulk substrate wafer, on which we perform homoepitaxy of AlN and then insert a 20 nm coherently strained GaN QW to obtain a high conductivity 2DEG channel due to the polarization discontinuity.  The sheet resistance of the 2DEG channel is in the $\sim$ 250 $\Omega/\square$ range with the inhomogeneity indicated in Figure \ref{fig1}(c).  We subsequently use this epiwafer to fabricate XHEMTs.  Before describing the process, we discuss the details of the formation of the conductive 2DEG and how it is different from the conventional AlGaN/GaN geometry used today.

Figure \ref{fig1}(d) shows the cross section of conventional AlGaN/GaN HEMTs developed on Si primarily for power electronics \cite{GaN_on_Si_Review}, and Figure \ref{fig1}(e) shows that of AlGaN/GaN HEMTs on SiC used for radio frequency (RF) power amplifiers \cite{HRL_Highspeed, Wolfspeed_SunkenFP}. HEMTs on silicon offer a low-cost option, but do require a thick GaN epitaxial layer to obtain high GaN crystalline quality by reduction of dislocation density by annihilation, and to keep the narrower bandgap silicon far from high electric field regions.  The nearly 17\% lattice mismatch between GaN and Si leads to wafer curvature, cracking, and defect formation, which require several buffer layers and defect mitigation schemes developed in the past decade \cite{Umeno_GaN_on_Si_Growth, Egawa_GaN_Buffer_On_Si}. The undesired mobile n-type carriers due to intentional shallow donor impurities in the thick GaN buffer layer is compensated by Fe or C impurity doping that introduce deep levels inside the energy bandgap of GaN \cite{Kuball_BufferDesign, Meneghini_Buffer_Trap}.  The high defect density and thermal boundary resistance at the nucleation layer/silicon interface introduce a bottleneck in heat dissipation \cite{Kuball_GaN_Substrate_TBR}. Although not explicitly shown in Figure \ref{fig1}, sapphire is another commonly used low-cost substrate, exhibiting similar issues due to its $\sim$ 14\% lattice mismatch with GaN and its even-lower-than-Si thermal conductivity.

Because the lattice mismatch between GaN and SiC is 3.4\% and SiC has a high thermal conductivity and larger bandgap than silicon, somewhat thinner epitaxial layers and better heat dissipation is obtained in AlGaN/GaN HEMTs on SiC.  The thinner epitaxial layers cause lower wafer curvature, though the dislocation density is still typically $\sim 10^9$/cm$^2$ even in the best GaN HEMTs on SiC that are currently favored for high power microwave power amplifiers. The thermal boundary resistance between the nitride and SiC layer remains a concern \cite{Kuball_Nucleation_TBR}.

Figure \ref{fig1}(f) shows the layer structure of the XHEMT on single-crystal AlN. The AlN homoepitaxial layer grown on AlN eliminates three things: 1) dislocations due to coherent epitaxy \cite{CrystalIS_AlNProgress}, 2) thermal boundary resistance at the substrate \cite{Gustavo_AlN}, and 3) the need for deep level compensation dopants because undoped AlN has ultrahigh resistivity and low RF loss \cite{EK_N-polar}.  The epitaxial layer thickness is reduced by 2X to 10X compared to growths on silicon or SiC. 

The choice of AlN substrates not only affects the crystal quality and the electrical and thermal properties of the transistor, but also drastically alters the charge distribution within the heterostructure. Just as the polarization discontinuity at the Al(Ga)N/GaN interface induces the desired 2DEG, the polarization charge at the AlN/GaN interface, now with the opposite sign, induces 2DHG as reported in our earlier studies \cite{Chuan_Hole_Oscillation, Reet_2DHG} . Therefore, unlike the conventional HEMT structures shown in Figure \ref{fig1}(d) and \ref{fig1}(e), the 2DEG and 2DHG are expected to coexist in the thin GaN channel layer of the AlN XHEMT structure. In this study, we find that the 2DHG, located approximately 18 nm below the electron channel, causes severe charge trapping effects during RF operation and limit the transistor's output power and efficiency. We further show that inserting a sheet of silicon donors can effectively compensate the 2DHG and eliminate its adverse effects on RF performance while simultaneously boosting the 2DEG conductivity.

\section{Results and Discussion}
\subsection{Heterostructure design and electrical transport characteristics}

\begin{figure}[t!]
\centering
\includegraphics[width=\textwidth]{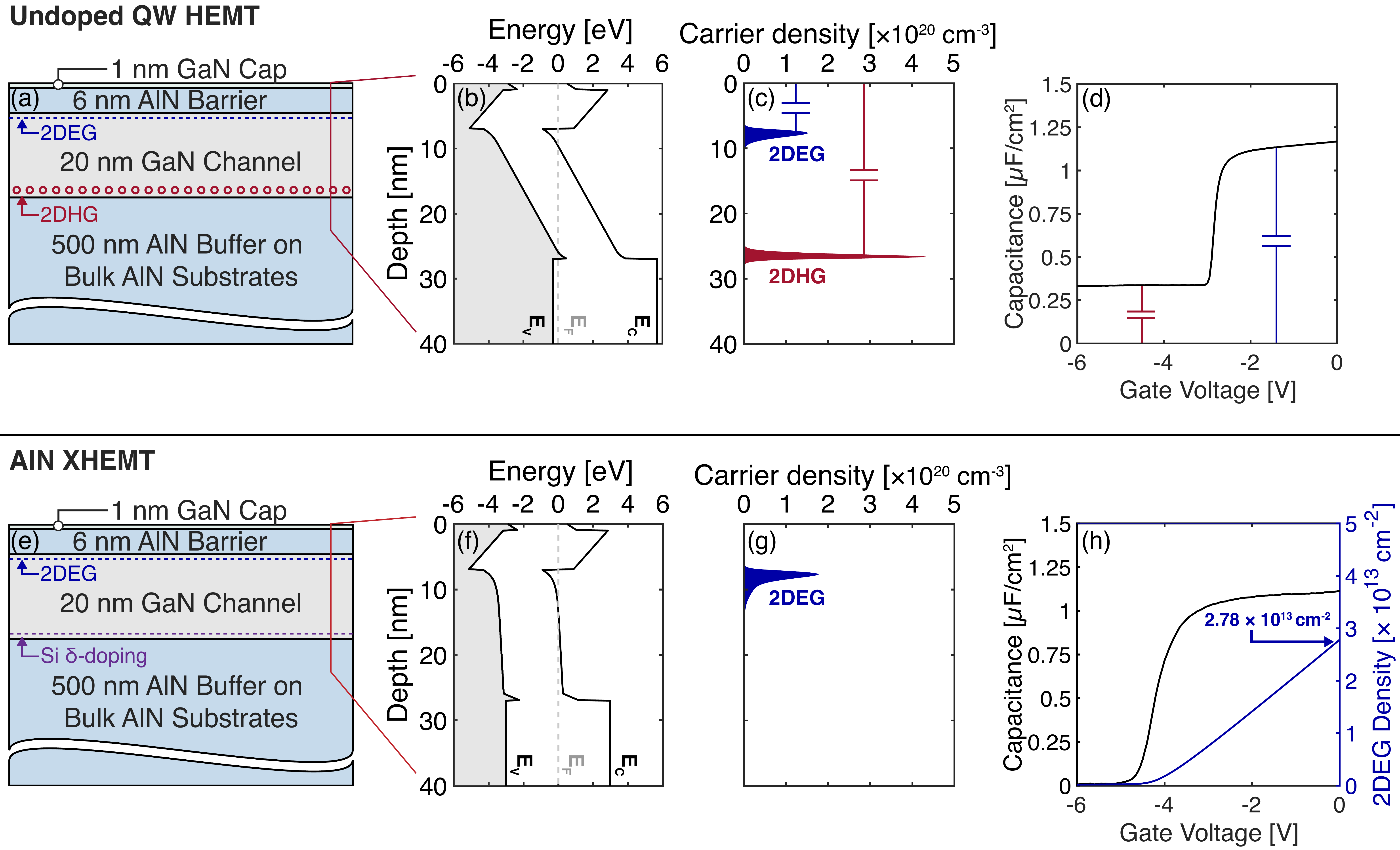}
\caption{Undoped QW HEMT on AlN (top row) vs AlN XHEMT (bottom row). Layer structure of the (a) undoped QW HEMT and (e) AlN XHEMT in this study, both consisting of epitaxial AlN/GaN/AlN layers on bulk AlN. AlN XHEMT uniquely refers to the structure with silicon $\delta$-doping thus only the electron channel is present. Otherwise, the two structures are nominally the same. Simulated energy band diagrams of (b) undoped QW HEMT and (f) AlN XHEMT. Silicon donors compensate for the net negative polarization charge at the bottom GaN/AlN interface on metal-polar AlN, shifting the Fermi level in GaN away from the valence band and closer to the midgap of AlN back barrier in AlN XHEMT. This results in a reduction of the average vertical electric field in GaN experienced by the 2DEG. Simulated carrier density profile of (c) undoped QW HEMT and (g) AlN XHEMT. Silicon $\delta$-doping in AlN XHEMT prevents accumulation of 2DHG at the bottom GaN/AlN interface and increases the 2DEG density in the channel by modulation doping effect. (d) C-V characteristics of the undoped QW HEMT. Two plateaus are observed, corresponding to the 2DEG and the 2DHG present in the device structure as indicated by the color-coded capacitor symbols, respectively. (h) C-V characteristics of the AlN XHEMT, confirming single electron channel operation.}
\label{fig2}
\end{figure}

Silicon $\delta$-doped and undoped AlN XHEMTs (hereafter referred to as AlN XHEMTs and undoped QW HEMTs, respectively) are compared in Figure \ref{fig2} in terms of their layer structures, equilibrium energy band diagrams and charge distributions, and capacitance–voltage characteristics. Figure \ref{fig2}(a) illustrates the epitaxial heterostructure of undoped QW HEMTs. The structure consists of a 500 nm homoepitaxial AlN buffer layer grown on single-crystal metal-polar AlN substrates from Asahi-Kasei Corporation \cite{DUV_Laser, 2-inch_AlN}, a 20 nm coherently strained GaN channel, and a 6 nm AlN top barrier capped with a 1 nm GaN layer. Figure \ref{fig2}(b) and \ref{fig2}(c) show the simulated energy band diagram and carrier density profile, respectively. In addition to the polarization-induced 2DEG at the top AlN/GaN interface, the calculations suggest that net negative polarization bound charges at the bottom GaN/AlN interface induce positive charges with a large density of $p_\text{s} \sim 3.9 \times 10^{13}$ cm$^{-2}$, confined by the valence band offset between GaN and AlN. We confirm the existence of the positive charges at this GaN/AlN interface by the capacitance-voltage (C-V) measurement performed on on-wafer circular Schottky diodes. As shown in Figure \ref{fig2}(d), a second capacitance plateau is observed beyond the 2DEG depletion, corresponding to a mobile sheet charge situated near the bottom GaN/AlN interface, indicated by the capacitor symbol in red. The capacitance at this second plateau is 0.34 $\mu$F/cm$^2$, corresponding to a depth of 26 nm below the surface where the positive charges are expected based on the simulation results. We attribute this second plateau to the presence of a polarization-induced two-dimensional hole gas (2DHG), corroborated by our earlier observation of a high-density 2DHG in an undoped GaN/AlN heterostructure grown on the single-crystal AlN substrates, where the growth was terminated after growing the GaN channel layer \cite{ZZ_2DHG, Reet_2DHG}. The influence of the 2DHG on electrical characteristics of undoped QW HEMTs remained largely unexplored to date.

To investigate the effect of the uncompensated positive charges at the GaN/AlN interface, we performed identical growths, differing only by the introduction of silicon $\delta$-doping with a donor density of 5 $\times$ $10^{13}$ cm$^{-2}$, inserted 1 nm above the bottom GaN/AlN interface, as shown in Figure \ref{fig2}(e). In this AlN XHEMT structure, the additional electrons supplied by the silicon donors compensate for the positive charges, shifting the valence band away from the Fermi level. The compensation provided by the silicon donors reduces the polarization effect at the GaN/AlN interface, thereby lowering the vertical electric field in the GaN QW and pushing the centroid of the electron wave function further away from the interface. This redistribution of electron wave function results in reduced interface roughness scattering, enhancing electron mobility. The remaining electrons, after compensation, increases the 2DEG density through the modulation doping effect, as they migrate to the opposite end of the QW and become confined. Hence, a concurrent increase in both electron mobility and 2DEG density is expected with silicon $\delta$-doping. The simulated energy band diagram and carrier density profile, shown in Figure \ref{fig2}(f) and \ref{fig2}(g), show the anticipated suppression of positive charges and the reduction of the average vertical field in the GaN QW. The carrier density-weighted average field in GaN QW is 2.84 MV/cm in AlN XHEMTs and 3.80 MV/cm in undoped QW HEMTs, respectively, calculated as $F_{\text{avg}} = \frac{\int n(z)F(z) dz}{\int n(z) dz}$, where $n(z)$ is the local electron density and $F(z)$ is the local electric field along the direction $z$ perpendicular to the sample surface. The C-V characteristics of an AlN XHEMT, as shown in Figure \ref{fig2}(h), confirm the charge compensation, showing a C-V profile closely aligned with the expected behavior of a GaN HEMT with a single electron channel with no 2DHG: the capacitance sharply decreases at the threshold voltage as the 2DEG is depleted and no additional plateau is observed.

Hall-effect measurements at room temperature performed on as-grown samples revealed a 2DEG of density and electron mobility of 1.80 $\times$ 10$^{13}$ cm$^{-2}$ and 717 cm$^2$/V$\cdot$s, corresponding to a sheet resistance of 485 $\Omega/\square$ in undoped QW HEMTs, and 3.16 $\times$ 10$^{13}$ cm$^{-2}$ and 644 cm$^2$/V$\cdot$s, corresponding to a sheet resistance of 307 $\Omega/\square$ in AlN XHEMT, respectively. The increase in the 2DEG density in the AlN XHEMT structure, owing to silicon $\delta$-doping, leads to a $\sim$ 36.8$\%$ reduction in sheet resistance. The 2DEG density of AlN XHEMTs at zero gate bias, extracted from the C-V curve shown in Figure \ref{fig2}(h), is 2.78 $\times$ 10$^{13}$ cm$^{-2}$. This slightly lower value compared to the Hall data is due partly to metal gate depletion and partly to surface fluorination during recess etching of silicon nitride. In a more recently grown AlN XHEMT sample with further growth optimization, although devices have not yet been fabricated, a 2DEG density of 3.21 $\times$ 10$^{13}$ cm$^{-2}$ and electron mobility of 855 cm$^2$/V$\cdot$s, corresponding to a sheet resistance of 227 $\Omega/\square$ were measured, resulting in a $\sim$ 53.1$\%$ reduction in sheet resistance compared to the undoped QW HEMTs \cite{Cindy_XHEMT}.

\subsection{Device design, DC performance, and small-signal RF characteristics of AlN XHEMTs}

\begin{figure}[h!]
\centering
\includegraphics[width=\textwidth]{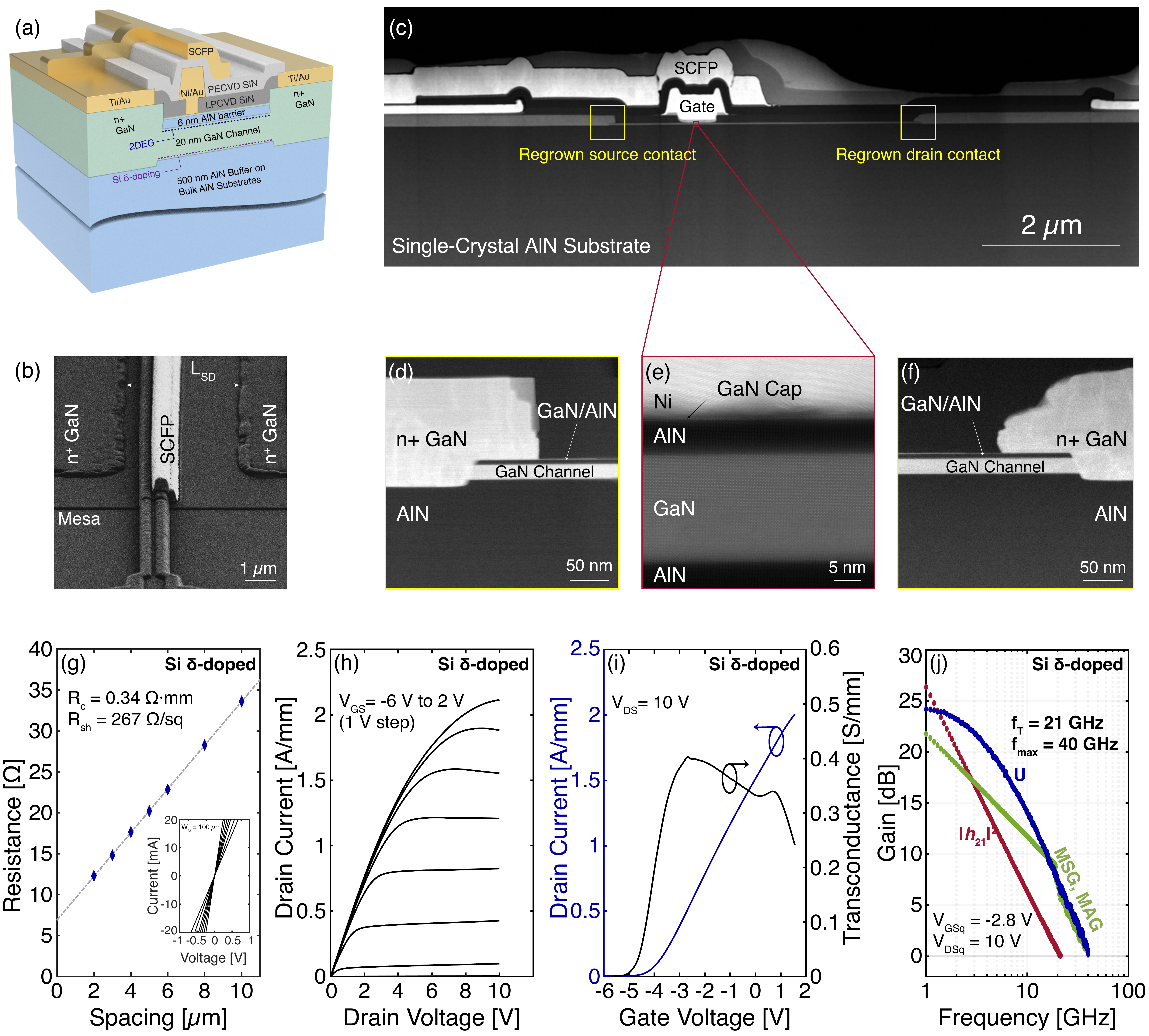}
\caption{AlN XHEMT. (a) Three-dimensional representation and (b) SEM image of a fully fabricated AlN XHEMT. (c) Cross-sectional STEM image showing the regrown source and drain contacts, a T-shaped gate, and a source-connected field plate. The cross-sectional images were taken along the plane that intersects the bridge connecting the field plate to the source pad. (d) Magnified STEM image showing the regrown GaN/2DEG interface on the source side. (e) Atomic-resoluation HAADF-STEM image revealing a sharp heterojunction with a coherently strained GaN channel layer under the gate. (f) Magnified STEM image showing the regrown GaN/2DEG interface on the drain side. Threading dislocations are widely observed within the relaxed regrown n+ GaN, but not seen in the pseudomorphically strained GaN channel of the AlN XHEMT. (g) Linear TLM analysis performed on AlN XHEMTs, showing a contact resistance $R_\text{c}$ = 0.34 $\Omega\cdot$mm between the ohmic metal and the 2DEG and a 2DEG sheet resistance $R_{\text{sh}}$ = 267 $\Omega/\square$. (h) Linear plot showing the family of I-V curves for an AlN XHEMT with a gate-to-source voltage ranging from -6 V to 2 V in steps of 1 V. (i) Drain current (blue line) and transconductance (black line) of an AlN XHEMT as a function of gate-to-source voltage, operating at a drain-to-source voltage of 10 V. (j) Semi-log plot showing the small-signal current gain (red line), unilateral gain (blue line), and maximum stable and available gain (green line). An AlN XHEMT biased at a gate-to-source voltage $V_{\text{GSq}}$ = -2.8 V and a drain-to-source voltage $V_{\text{DSq}}$ = 10 V revealed $f_\text{T}$/$f_{\text{max}}$ = 21/40 GHz. The extra parasitic delays introduced by device probe pads were not de-embedded. The dimensions of the measured AlN XHEMTs are L$_\text{G}$ = 0.45 $\mu$m, L$_{\text{SD}}$ = 4.5 $\mu$m, L$_{\text{GD}}$ = 3.05 $\mu$m, and W$_\text{G}$ = 2 $\times$ 100 $\mu$m.}
\label{fig3}
\end{figure}

Fully fabricated AlN XHEMTs are shown schematically in Figure \ref{fig3}(a). The source and drain ohmic contacts were formed by regrowing a heavily silicon doped n+ GaN layer, followed by an inductively coupled plasma (ICP) etch extending approximately 10 nm into the AlN buffer layer to expose the 2DEG sidewall. The devices were passivated with a 106-nm thick near stoichiometric silicon nitride (SiN$_\text{x}$) layer deposited by low pressure chemical vapor deposition (LPCVD). A nickel/gold metal stack was used to form the Schottky gate contact, and a source-connected field plate (SCFP) was implemented, separated from the gate metal by a 118 nm thick silicon nitride layer deposited by plasma-enhanced chemical vapor deposition (PECVD). A scanning electron microscope (SEM) image of a fully processed AlN XHEMT, taken at a 70-degree angle, is shown in Figure \ref{fig3}(b). The undoped QW HEMTs were fabricated using the same process flow. Details of the device fabrication can be found in the experimental section.

The high-resolution cross-sectional scanning transmission electron microscopy (STEM) image (Figure \ref{fig3}(c)) shows the device cross-section, featuring a gate length (L$_{\text{G}}$) of 0.45 $\mu$m, a source-to-drain distance (L$_{\text{SD}}$) of 4.5 $\mu$m, and a gate-to-drain distance (L$_{\text{GD}}$) of 3.05 $\mu$m. As depicted in the device schematic in Figure \ref{fig3}(a), SCFPs are connected to the source pad via a 5-$\mu$m-wide bridge covering the access region over a device width (W$_{\text{G}}$) of 100 $\mu$m to minimize parasitic gate-to-source capacitance. Cross-sectional images were taken along the plane intersecting this bridge. Figure \ref{fig3}(d) and \ref{fig3}(f) show the source and drain ohmic contacts, respectively, formed with a regrown n++ GaN layer. To ensure intimate regrown contacts, the ICP etch parameters were selected to expose the 2DEG sidewall at a 50-degree angle with respect to the epitaxial interface, with the regrown n++ GaN layer extending 80 nm into the access region. The atomic-resolution STEM image in Figure \ref{fig3}(e), taken under the T-shaped gate, confirms that the 20-nm GaN channel remains coherently strained on single-crystal AlN substrates with atomically sharp interfaces. To facilitate easier viewing of atomic details, a larger version of the STEM image is provided in Supplementary Figure S1.

Following device fabrication, the contact resistance ($R_\text{c}$) and sheet resistance ($R_{\text{sh}}$) were extracted using the linear transfer length method (TLM), as shown in Figure \ref{fig3}(g). An average $R_\text{c}$ of 0.39 $\pm$ 0.04 $\Omega\cdot$mm and $R_{\text{sh}}$ of 276 $\pm$ 9.5 $\Omega/\square$ were obtained from multiple TLM measurements across the sample. The extracted $R_{\text{sh}}$ closely aligns with the $R_{\text{sh}}$ of 283 $\Omega/\square$ obtained by the Hall effect measurement on an on-wafer van der Pauw pattern post-device fabrication. The lower $R_{\text{sh}}$, compared to the value measured on the as-grown sample prior to device fabrication, is attributed to the increased 2DEG density due to surface passivation with LPCVD SiN$_\text{x}$.

Figure \ref{fig3}(h) shows the representative output characteristics of AlN XHEMTs with the nominally same dimensions with the device in Figure \ref{fig3}(c). At room temperature, the device exhibits a maximum drain current density exceeding 2 A/mm at a gate voltage of 2 V. This high drain current density is realized without aggressive lateral scaling of the device, owing to the enhanced 2DEG density provided by the AlN barrier and silicon $\delta$-doping. Figure \ref{fig3}(i) shows the transfer characteristics of the AlN XHEMTs. At a fixed drain voltage of 10 V, a threshold voltage of -4.2 V and a peak extrinsic transconductance exceeding 0.4 S/mm were extracted. Figure \ref{fig3}(j) shows the small-signal unilateral gain ($U$), current gain ($|h_{\text{21}}|^2$), maximum stable gain ($MSG$), and maximum available gain ($MAG$), which were extracted from the measured S-parameters of a device biased at a gate voltage of -2.8 V and a drain voltage of 10 V. The extra parasitic delays introduced by device probe pads were not de-embedded. The cut-off frequency ($f_\text{T}$) and maximum oscillation frequency ($f_{\text{max}}$) of 21 and 40 GHz, respectively, were extracted from $|h_{\text{21}}|^2$ and $U$, both of which exhibited the expected -20 dB/dec slope. The DC and small-signal characteristics of the undoped QW HEMTs with the same device dimensions are summarized in Supplementary Figure S2 for comparison.

\subsection{Large-signal RF characteristics of AlN XHEMTs}

\begin{figure}[t!]
\centering
\includegraphics[width=\textwidth]{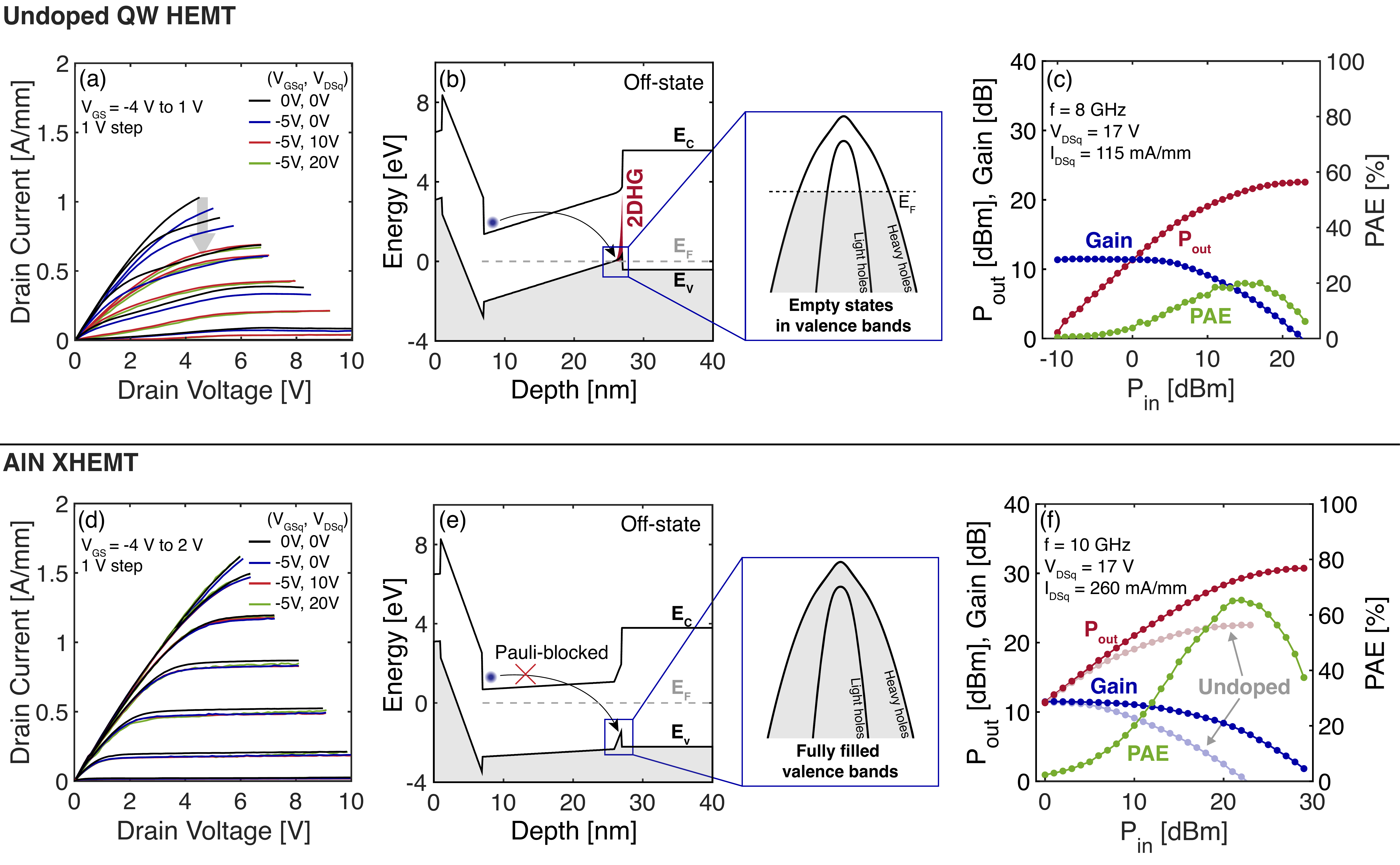}
\caption{Undoped QW HEMT on AlN (top row) vs AlN XHEMT (bottom row), both fabricated using the same process flow. (a) and (d) Pulsed $I_\text{D}-V_{\text{DS}}$ characteristics. Cold bias condition ($V_{\text{GSq}}$/$V_{\text{DSq}}$ = 0 V/0 V) was used as a reference, and the applied stress bias conditions were $V_{\text{GSq}}$/$V_{\text{DSq}}$ = -5V/0V, -5V/10V, and -5V/20V for both undoped QW HEMTs and AlN XHEMTs. (b) and (e) Simulated energy band diagrams under the gate in the off-state. The undoped QW HEMTs exhibited severe current collapse. We hypothesize when the transistor switches from on to off state, some electrons in the channel flow into the empty states in the valence band - i.e. holes - at the bottom GaN/AlN interface; however, these "trapped" electrons takes ms or longer to be fully released when the transistor switches back on, which leads to reduction in the mobile electrons in the channel therefore reduced drain current under pulsed bias conditions. On the other hand, AlN XHEMTs show negligible current collapse since these are no empty states in the valence band thanks to the electrons from the silicon donors. (c) and (f) CW load-pull power performance measurement as a function of input power, tuned for maximum PAE. In the undoped QW HEMTs, the maximum $P_{\text{out}}$ and peak PAE are limited to 0.90 W/mm and 20$\%$, respectively, due to severe gain compression caused by electron trapping. In contrast, AlN XHEMTs exhibit a much higher maximum $P_{\text{out}}$ of 5.92 W/mm and a peak PAE of 65$\%$.} 

\label{fig4}
\end{figure}

In GaN-based HEMTs, devices exhibiting excellent DC and small-signal RF performance often fail to achieve high RF output power due to severe charge trapping effects, which lead to current collapse under RF operation conditions \cite{NRL_Trapping, Vetury_Trapping}. Therefore, the extent and origins of charge trapping effects in newly designed heterostructures for HEMTs need to be carefully evaluated. A quick measurement to discern charge trapping is to observe whether the transistor channel current recovers fully or not when switching from off-state to on-state: charge trapping in any part of a transistor will prevent the transistor drain current to recover fully within a time period shorter than the charge de-trapping time. The ultimate test is to measure the transistor amplification performance as a function of the RF input signal power so that the entire usable I-V area is accessed to amplify the RF signal: a transistor with minimal trapping should exhibit RF performance consistent with that predicted from the transistor DC I-V characteristics.

To this end, pulsed current-voltage ($I_\text{D}-V_{\text{DS}}$) measurements were applied to both the undoped QW HEMTs and the AlN XHEMTs using 500 nm long pulses at a 1 ms period to investigate their dynamic behavior under large-signal drive conditions. As shown in Figure \ref{fig4} (a), the undoped QW HEMTs show severe current collapse exceeding 30$\%$ under a stress bias condition of $V_{\text{GSq}}$, $V_{\text{DSq}}$ = -5 V, 20 V, compared to the cold bias condition of $V_{\text{GSq}}$, $V_{\text{DSq}}$ = 0 V, 0 V. In contrast, a dramatic improvement in trapping phenomena was achieved in XHEMTs with silicon $\delta$-doping. The AlN XHEMTs demonstrated negligible current collapse under the same stress bias condition of $V_{\text{GSq}}$, $V_{\text{DSq}}$ = -5 V, 20 V, as shown in Figure \ref{fig4} (d).

The large current collapse, consistently observed in undoped QW HEMTs with varying dimensions across multiple samples that were grown and processed independently, is believed to be in part caused by hot electrons being captured in empty states in the valence band at the bottom GaN/AlN interface, located approximately 20 nm below the 2DEG channel, as shown in the simulated off-state energy band diagram in Figure \ref{fig4} (b). Once captured, these electrons are unable to return immediately to their original state when the transistor transitions back to its on-state during the positive cycle of the gate signal \cite{Wenshen_CompoundSemiconductor}, leading to partial depletion of the 2DEG. In the case of AlN XHEMTs, silicon $\delta$-doping shifts the Fermi level closer to the midgap, effectively filling the valence band states. As a result, the capture of electrons is Pauli-blocked, as shown in the off-state energy band diagram of AlN XHEMTs in Figure \ref{fig4}(e), preventing partial depletion of 2DEG.

The continuous-wave (CW) large-signal performance of the undoped QW HEMTs was investigated to confirm the charge trapping effects. The undoped QW HEMT, biased at $V_{\text{DSq}}$ = 17 V and $I_{\text{DSq}}$ = 0.115 A/mm, demonstrated a maximum output power density ($P_{\text{out}}$) and a peak power-added efficiency (PAE) limited to 0.90 W/mm and 20$\%$, respectively, at 8 GHz. These values fall far short of the expected $P_{\text{out}}$ based on its output characteristics and bias point. As shown in Figure \ref{fig4}(c), the $P_{\text{out}}$ and PAE of undoped AlN HEMTs are constrained by the gain compression, caused by electron trapping, as anticipated from the previous pulsed $I_\text{D}-V_{\text{DS}}$ measurements.

The CW large-signal measurements performed on AlN XHEMTs further confirmed the effectiveness of silicon $\delta$-doping in suppressing RF dispersion. Biased at $V_{\text{DSq}}$ = 17 V and $I_{\text{DSq}}$ = 0.26 A/mm, AlN XHEMTs exhibited a maximum $P_{\text{out}}$ of 5.92 W/mm and a peak PAE of 65$\%$, as shown in Figure \ref{fig4}(f) when tuned for maximum PAE; under the same matching conditions, the associated $P_{\text{out}}$ at the peak PAE was 4.2 W/mm. These results, enabled by silicon $\delta$-doping, represent nearly 6-fold and 3-fold increases in $P_{\text{out}}$ and PAE, respectively, compared to the undoped QW HEMTs. Further increase in $P_{\text{out}}$ is currently limited by the device breakdown voltage due to a non-optimized electric field management near the gate edge.

In previous studies, we reported that AlN/GaN/AlN HEMTs on bulk AlN substrates with a thick, relaxed 250-nm GaN layer exhibit low RF dispersion \cite{EK_DRC, Cindy_QW}. This low dispersion is presently attributed to the large separation between the electron channel at the top AlN/GaN interface and empty states --- if present and not compensated by other defects --- in the valence band at the bottom GaN/AlN interface by the thick GaN layer. Furthermore, the thick GaN layer likely possesses compensating defects, which also hinder electrons from being captured in the empty states. This highlights a challenge in scaling the GaN channel thickness on AlN - the most suitable back barrier in terms of its high energy barrier to confine electrons, its atomically sharp isostructural interface with GaN, and its high thermal conductance. While increasing the channel thickness mitigates RF current collapse, a high density threading dislocation density is inevitable in a thick, relaxed GaN channel layer, which will negate the key advantages offered by the single crystal AlN substrate illustrated in Figure \ref{fig1}. Given that the pseudomorphic GaN thickness is about 20 nm on AlN due to their lattice mismatch, and that the 2DHG can develop in a pseudomorphic GaN on AlN as thin as 3 nm, it is essential to introduce donor doping to compensate the negative sheet charge at the bottom of the GaN/AlN interface.

Clearly, many aspects of the AlN XHEMTs should be improved and explored, including further reduction of the GaN channel thickness and mobile electron concentration under the gate while improving electron mobility, further increase in breakdown voltage, studies of device reliability, device-processing-circuit co-design to maximize the advantages that AlN offers, etc. However, the excellent RF performance achieved in these early generations of the AlN XHEMTs represents the first success toward developing active RF devices based on single-crystal AlN.

\subsection{Benchmarking}

To evaluate the potential of AlN XHEMTs, we benchmark the large signal RF performance obtained on these first-generation XHEMTs grown on Al-polar AlN against their counterpart: metal-polar, single-channel GaN HEMTs reported in the literature.

\begin{figure}[h!]
\centering
\includegraphics[width=\textwidth]{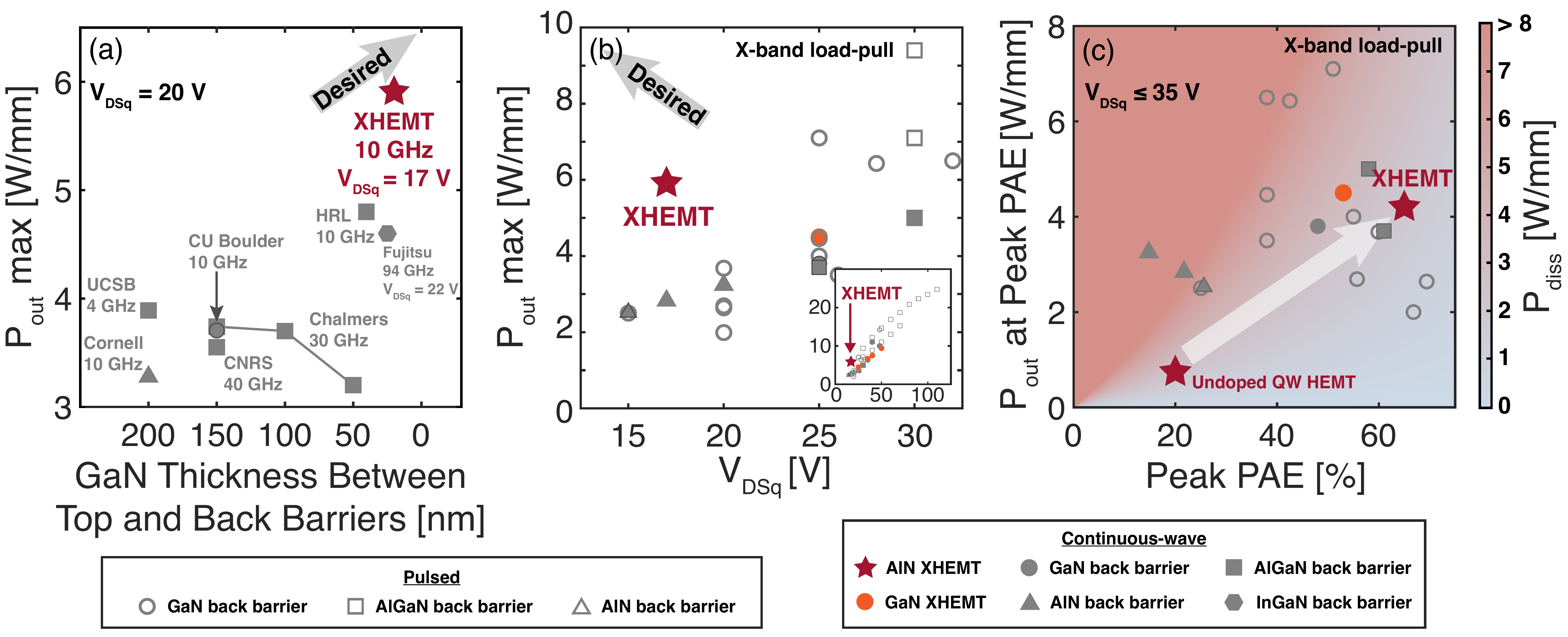}
\caption{Benchmarking AlN XHEMT with metal-polar, single-channel nitride HEMTs on large signal RF power amplification. Hollow symbols are for pulsed and solid symbols for CW performance. (a) Maximum $P_{\text{out}}$ measured at $V_{\text{DSq}}$ = 20 V and GaN channel thickness (GaN thickness between the top and back barriers), in comparison to HEMTs with an AlN (triangle), AlGaN (square), GaN (circle), and InGaN (hexagon) back barrier reported in the literature \cite{Chalmers_2019, CNRS_2023, UCSB_2013, Austin_FirstRF, InGaN_94GHz, HRL_10GHz, Schafer_2013}. Generally speaking, the conventional GaN back barrier is impurity doped with Fe or C, which possesses deep levels in GaN, and the GaN channel layer has a compensation doping concentration below $1 \times 10^{17}$ cm$^{-3}$; accordingly, the GaN channel thickness is between 100 and 200 nm. (b) $P_{\text{out}}$ as a function of $V_{\text{DSq}}$, comparison to previously reported HEMTs in X-band. The performance of GaN XHEMTs --- AlGaN/GaN HEMTs on single-crystal GaN substrates --- is marked in orange circle \cite{Sardin_2014, Lee_2005, GaN_on_Diamond, Wu_2004, Luo_2011, Peng_2011, Tao_2016, Resca_2014, Schafer_2013, Piotrowicz_2010, Piotrowicz_2008, GaN_XHEMT, HRL_2012, Fujitsu_first, Fujitsu_Record}. (c) Comparison of $P_{\text{out}}$ at peak PAE versus peak PAE in X-band with $V_{\text{DSq}}$ $\leq$ 35 V. AlN XHEMT shows significant improvements over the undoped QW HEMTs, moving closer to the desired upper-right corner. In the context of thermal properties, only $P_{\text{out}}$ values at peak PAE from Figure \ref{fig4} are shown in this benchmark plot. The color map shows the dissipated power ($P_{\text{diss}}$) at a given $P_{\text{out}}$ and PAE, derived under the assumption of 8 dB transducer power gain at peak PAE.}
\label{fig5}
\end{figure}

Figure \ref{fig5}(a) benchmarks the large-signal performance of silicon $\delta$-doped AlN XHEMTs in terms of $P_{\text{out}}$ as a function of GaN channel thickness (GaN thickness between the top and the bottom barrier). As the GaN channel thickness decreases, large-signal RF amplification becomes increasingly challenging due to factors such as mobility degradation and higher DC-RF dispersion \cite{Chalmers_2019}. By incorporating silicon $\delta$-doping into the AlN XHEMT structure to overcome these challenges, this work reports the first large-signal operation of double-heterostructure HEMTs with a channel thickness at or below 20 nm. Moreover, the reported $P_{\text{out}}$ is the highest achieved for $V_{\text{DSq}}$ at or below 20 V among all metal-polar, single-channel GaN HEMTs though not all available data in the literature are at 10 GHz.

In Figure \ref{fig5}(b), $P_{\text{out}}$ of metal-polar, single-channel GaN HEMTs, including but not limited to double-heterostructure HEMTs, is plotted against $V_{\text{DSq}}$ in the frequency range from 8 to 12 GHz. At a given $V_{\text{DSq}}$, AlN XHEMTs deliver significantly larger $P_{\text{out}}$ compared to conventional AlGaN/GaN HEMTs, owing to their higher 2DEG density. The values used to create the benchmark figure are summarized in the supplementary material. Lastly, Figure \ref{fig5}(c) shows the associated $P_{\text{out}}$ at peak PAE of AlN XHEMTs compared with those of GaN HEMTs in Figure \ref{fig5}(b), measured at $V_{\text{DSq}}$ at or under 35 V in the X-band. A dramatic improvement in the large-signal RF performance of AlN XHEMTs is highlighted here, compared to the undoped QW HEMT. When the problem of charge trapping thus RF dispersion is addressed in the AlN XHEMT, the HEMT performance on the AlN platform moves closer to the upper-right desired corner --- which is for simultaneously higher $P_{\text{out}}$ and PAE with a cooler device junction.

\section{Conclusion}
This study introduces AlN XHEMTs, leveraging a thin GaN channel sandwiched between AlN layers on single-crystal AlN substrates. The epitaxial layers in the AlN XHEMT structure exhibit nearly a million-fold reduction in dislocation density compared to the conventional HEMT structure grown on foreign substrates, as well as the highest reported thermal boundary conductance at the growth interface among the material systems used in GaN HEMTs. In addition, silicon $\delta$-doping, inserted at the bottom of the GaN QW, enhances the 2DEG conductivity and enables dispersion-free operation by Pauli-blocking electron capture at the GaN/AlN interface where net negative polarization-bound charges are present. Experimental results confirm the potential of silicon $\delta$-doped AlN XHEMTs, demonstrating a maximum output power density of 5.92 W/mm and a peak power added efficiency of 65$\%$ at 10 GHz under a drain bias of 17 V. While further optimization of contact resistance and device design --- particularly field management thus increasing the quiescent drain bias --- is essential to achieve higher output power, these findings represent a significant milestone in the development of efficient RF electronics on the AlN platform that promises excellent thermal management and thin device layers by epitaxial growth.

% Experimental section

\section{Experimental Section}
\threesubsection{Epitaxial growth and characterization of electrical transport properties.}\\
AlN XHEMT heterostructure was grown on single-crystal AlN substrates from Asahi-Kasei, with dislocation densities below $10^4$ cm$^{-2}$. Epitaxial growth was performed using a Veeco GEN10 plasma-assisted molecular beam epitaxy (MBE) system, equipped with Al, Ga and Si standard effusion cells for metal flux control, as well as an RF plasma source for active nitrogen gas supply. Film growth was monitored in situ using a KSA Instruments reflection high-energy electron diffraction (RHEED) apparatus with a Staib electron gun operating at 14.5 kV and 1.45 A. 

Following the cleaning process, a $\sim$ 500 nm AlN buffer layer was grown under metal-rich conditions at a thermocouple temperature of T$_\text{c}$ = 1040 $^\circ$C, achieving step-flow growth mode. Excess Al droplets were desorbed in-situ by raising the substrate temperature by 50 $^\circ$C, with the desorption process monitored via RHEED intensity. The substrate was then cooled down to T$_\text{c}$ $\sim$ 840 $^\circ$C for GaN channel growth.

The active region, consisting of a $\delta$-doped GaN channel layer, an AlN barrier, and a GaN cap, was grown continuously under metal-rich conditions without growth interruptions. An RF plasma power of 200 W and an N$_2$ gas flow of 0.35 sccm were maintained during active region growth, corresponding to a growth rate of 0.2 $\mu$m/hour. Silicon $\delta$-doping was incorporated after depositing 1 nm GaN. The $\delta$-doping condition was calibrated using a separate MBE-grown silicon-doped GaN sample, with a silicon cell temperature of 1300 $^\circ$C, yielding a 3D doping density of 4.6 $\times$ 10$^{19}$ cm$^{-3}$.

Upon completion of epitaxial growth, the substrate was cooled immediately to room temperature, and excess Ga droplets were removed ex situ using HCl. The growth of undoped QW HEMTs followed the same procedure as that of AlN XHEMTs, except without the Silicon $\delta$-doping step. Following Ga droplet removal, a Nanometrics Hall system was used to measure electron mobility and 2DEG density in the as-grown samples using soldered indium corner contacts to the 2DEG.

\threesubsection{Device fabrication.}\\
Device fabrication began with patterning the sample for source and drain ohmic contact definition. SiO$_2$ and Cr hard masks were blanket-deposited on a 1$\times$1 cm$^2$ AlN XHEMT sample via low-power PECVD and electron-beam evaporation, respectively. The sample was then patterned using photolithography. Cr in the open contact windows was removed via O$_2$/Cl$_2$ ICP etching, and SiO$_2$ was etched using CF$_4$/CHF$_3$ reactive ion etching (RIE), with the patterned Cr serving as a hard mask after the photoresist was removed. The 2DEG sidewall was subsequently exposed by BCl$_3$ ICP etch, which extended approximately 10 nm into the AlN back barrier. The Cr layer was removed using a ceric ammonium nitrate-based wet etchant, and the SiO$_2$ hard mask was laterally recessed into the access region by approximately 80 nm using a diluted buffered oxide etchant (BOE). The patterned sample was loaded into a molecular beam epitaxy chamber, where a 60-nm thick n+ GaN layer was regrown.

The regrown n+ GaN outside the source and drain regions was lifted off using BOE, and devices were mesa-isolated by a BCl$_3$ ICP etch that extended into the AlN back barrier. Surface-passivation was then performed by depositing 106-nm thick near stoichiometric SiN$_\text{x}$ in a LPCVD chamber at a thermocouple temperature of 750 $^{\circ}$C. Dichlorosilane and ammonia precursors were used to grow near stoichiometric SiN$_\text{x}$. Non-alloyed source and drain ohmic contacts were then metallized by patterning the sample via photolithography, followed by SiN$_\text{x}$ removal using CHF$_3$/O$_2$ RIE and electron-beam evaporation of Ti/Au = 40/100 nm.

Gate stems were defined by a 100 keV JEOL 6300 electron beam lithography (EBL) system, followed by a gate recess etch performed via low-power SF$_6$ ICP etch to form Schottky contacts on GaN. The head width of T-shaped gates was defined using the same EBL system, and gates were metallized by electron-beam evaporation of Ni/Au = 40/350 nm. For RF measurements, coplanar waveguide bonding pads were connected to the transistor electrodes through electron-beam evaporating Ti/Au = 40/360 nm on a sample patterned via photolithography. A 116 nm-thick SiN$_\text{x}$ layer was then blanket deposited using PECVD. The sample was subsequently patterned by photolithography and SiN$_\text{x}$ on the transistor electrodes was removed via CHF$_3$/O$_2$ RIE to facilitate probing for electrical measurements. Lastly, SCFPs were defined using EBL and metallized by electron-beam evaporation of Ti/Au = 40/400 nm.

\threesubsection{DC and RF characterization}\\
The capacitance-voltage, and pulsed current-voltage characteristics of the HEMTs were measured using a Cascade Microtech Summit 11000 probe system and a Keithley 4200A-SCS parameter analyzer. Small-signal RF characterization was performed by measuring scattering using an Agilent E8364B vector network analyzer, with the DC bias supplied by an Agilent 4156C parameter analyzer. The transfer and output characteristics of the HEMTs were measured using the same system. The measurements were calibrated using short, open, load, and through impedance standards with Infinity ground-signal-ground (GSG) probes. The large-signal RF characterization was performed using a Maury Microwave MT2000 mixed signal active load-pull system with Infinity GSG probes. The optimum load and source reflection coefficients, tuned for maximum PAE, were $\Gamma_\text{L}$ = 0.23+0.32$i$ and $\Gamma_\text{S}$ = -0.68+0.37$i$, respectively, for the AlN XHEMT, and $\Gamma_\text{L}$ = 0.72+0.35$i$ and $\Gamma_\text{S}$ = -0.69+0.35$i$, respectively, for the undoped QW HEMT.

\threesubsection{Electron microscopy}\\
A cross-section lamella was prepared using the Thermo Fisher Helios G4 UX Focused Ion Beam. Protective C and Pt layers were deposited on the lamella and prepared with a final milling step of 5 keV to reduce damage. Scanning transmission electron microscopy (STEM) measurements were taken with an aberration-corrected Thermo Fisher Spectra 300 CFEG operated at 300 keV.

% Acknowledgements
\medskip
\textbf{Acknowledgements} \par %delete if not applicable))
This work was supported in part by ARO (device conceptualization, epitaxy, demonstration and characterization), under Grant No. W911NF-22-2-0177, by DARPA THREADS program (X-band device fabrication and characterization), Asahi-Kasei Corporation (substrates and epitaxy), and performed at the Cornell Nanoscale Facility, an NNCI member supported by NSF Grant No. NNCI-2025233. This work made use of the electron microscopy facility of the Cornell Center for Materials Research (CCMR) with support from the National Science Foundation Materials Research Science and Engineering Centers (MRSEC) program (DMR1719875). The Thermo Fisher Spectra 300 X-CFEG was acquired with support from PARADIM, an NSF MIP (DMR-2039380) and Cornell University. E.K. and N.P. acknowledge support from National Science Foundation Graduate Research Fellowship under Grant No. DGE2139899.
% References
\medskip

\clearpage
\setcounter{figure}{0}
\textbf{Supporting Information}
\begin{figure}[H]
\renewcommand{\thefigure}{S\arabic{figure}}
	\centering
        \includegraphics[width=\textwidth]{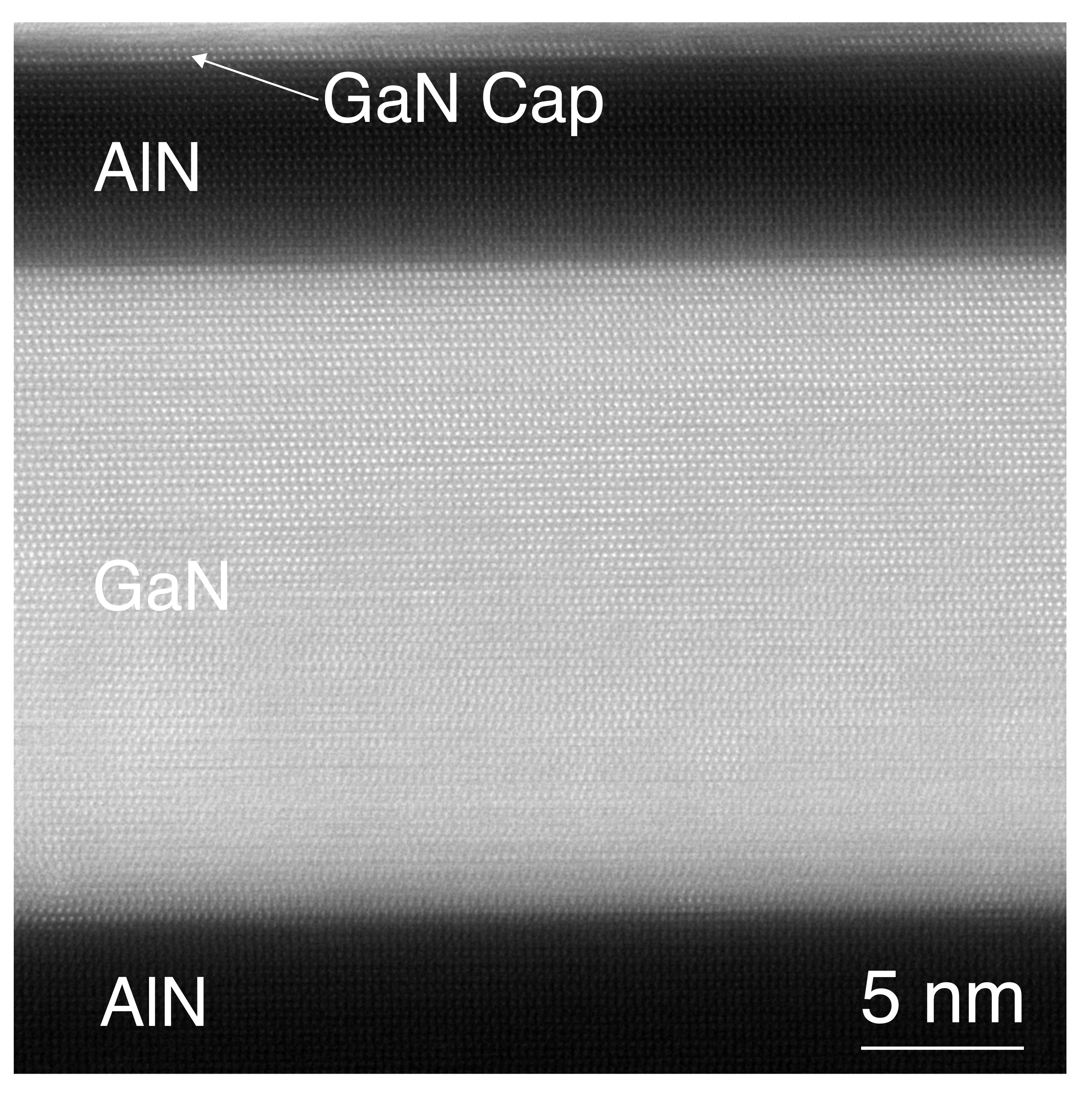}
	\caption{Atomic-resolution HAADF-STEM image, taken under the Ni Schottky gate of a fully fabricated device, revealing sharp interfaces in the GaN/AlN/GaN/AlN heterostructure and a coherently strained 20 nm GaN channel layer.}
	\label{figS1}
\end{figure}

\begin{figure}
\renewcommand{\thefigure}{S\arabic{figure}}
	\centering
        \includegraphics[width=\textwidth]{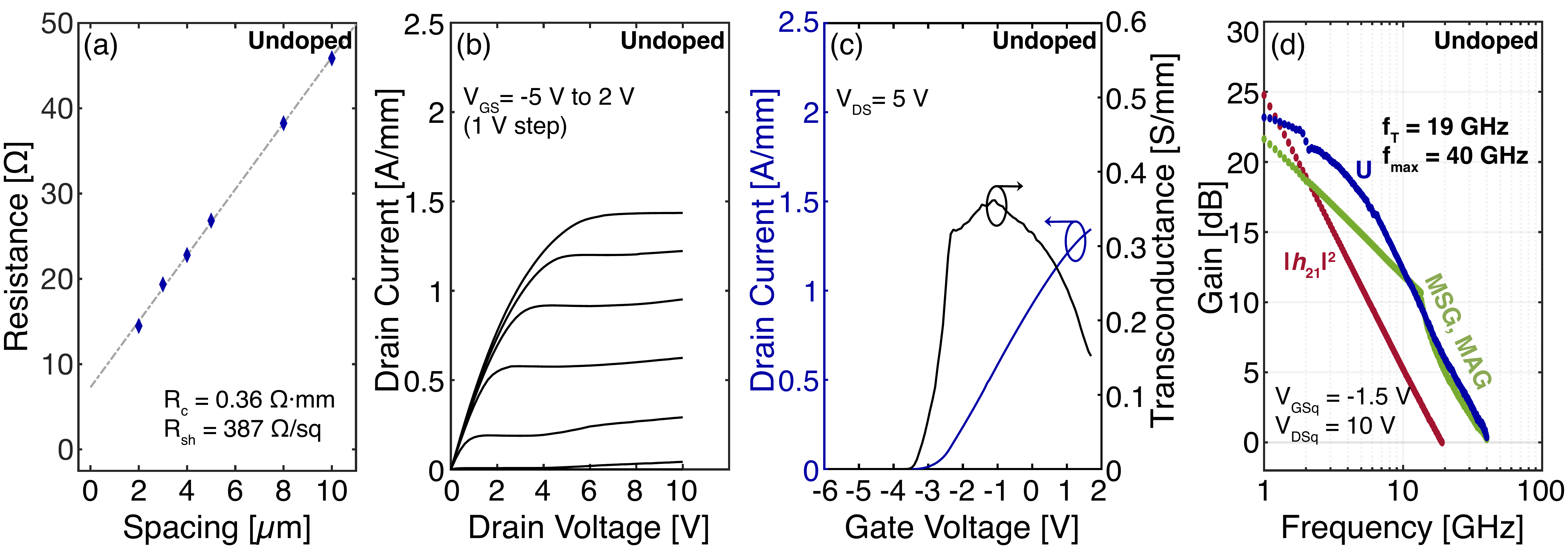}
	\caption{DC and small-signal characteristics of undoped QW HEMTs, fabricated using the same process flow as AlN XHEMTs. (a) Linear TLM analysis, showing a contact resistance $R_\text{c}$ = 0.36 $\Omega\cdot$mm between the ohmic metal and the
2DEG and a 2DEG sheet resistance $R_{\text{sh}}$ = 387 $\Omega/\square$. (b) Linear plot showing the family of I-V curves with gate-to-source voltage ranging from -5 V to 2 V in 1 V steps. (c) Drain current (blue line) and extrinsic transconductance (black line) as a function of gate-to-source voltage, measured at a drain-to-source voltage of 5 V. The undoped QW HEMTs and AlN XHEMTs typically exhibit on/off current ratios of 6 and 4 orders of magnitude, and a hard 3-terminal breakdown voltages $>$ 200 V and 35 V, respectively. Further optimization will be needed to improve electric field management in AlN XHEMTs to increase the breakdown voltage. (d) Semi-log plot showing the small-signal current gain (red line), unilateral gain (blue line), and maximum stable and available gain (green line). An undoped QW HEMT biased at a gate-to-source voltage $V_{\text{GSq}}$ = -1.5 V and a drain-to-source voltage $V_{\text{DSq}}$ = 10 V revealed a peak $g_\text{m}$ of 0.36 S/mm and $f_\text{T}$/$f_{\text{max}}$ = 19/40 GHz, less than 10\% lower than that of AlN XHEMTs. The dimensions of the measured undoped QW HEMTs were designed to be identical to the AlN XHEMTs shown in Figure \ref{fig3}: L$_\text{G}$ = 0.45 $\mu$m, L$_{\text{SD}}$ = 4.5 $\mu$m, L$_{\text{GD}}$ = 3.05 $\mu$m, and W$_\text{G}$ = 2 $\times$ 100 $\mu$m.}
	\label{figS1}
\end{figure}

\clearpage
\begin{table}
\renewcommand{\thetable}{S\arabic{table}}
\centering
\label{table1} 

\clearpage
\begin{tabular}{ |c|c|c|c|c|c|c| } 
 \hline
 GaN thickness & $V_{\text{DSq}}$ & $P_{\text{out}}$ & Frequency & Back barrier & Institute & Reference \\ 
 between barriers [nm] & [V] & [W/mm] & [GHz] & & &  \\
 \hline
 50 & 20 & 3.2 & 30 & AlGaN & Chalmers & \cite{Chalmers_2019}  \\ 
 100 & & 3.7 & & & &  \\
 150 & & 3.74 & & & &  \\
 \hline
 150 & 20 & 3.55 & 40 & AlGaN & CNRS & \cite{CNRS_2023}  \\ 
 \hline
 200 & 20 & 3.89 & 4 & AlGaN & UCSB & \cite{UCSB_2013}  \\ 
 \hline
 200 & 20 & 3.3 & 10 & AlN & Cornell & \cite{Austin_FirstRF}  \\ 
 \hline
 25 & 22 & 4.6 & 94 & InGaN & Fujitsu & \cite{InGaN_94GHz}  \\ 
 \hline
 40 & 20 & 4.8 & 10 & AlGaN & HRL & \cite{HRL_10GHz}  \\ 
 \hline
 150 & 20 & 3.68 & 10 & GaN & CU Boulder  & \cite{Schafer_2013}  \\ 
 \hline
\end{tabular}
\caption{Previously reported output power density and power-added efficiency of various GaN HEMTs with different GaN thicknesses between the top and bottom barriers. These values were used to create Figure 5(a).}
\end{table}

\clearpage
\newpage
\begin{table} 
\renewcommand{\thetable}{S\arabic{table}}
	\centering
	\label{table2}
\begin{tabular}{ |c|c|c|c|c|c|c| } 
 \hline
 Configuration & Back barrier & $V_{\text{DSq}}$ & $P_{\text{out}}$ & PAE & Institute & Reference \\ 
  & & [V] & [W/mm] & [\%] &  &   \\ 
 \hline
 CW & GaN & 25 & 3.8 & 48 & CU Boulder & \cite{Sardin_2014}\\
  & &  35 & 6.35 & 50 & TriQuint & \cite{Lee_2005}\\
  & &  40 & 11 & 51 & BAE systems & \cite{GaN_on_Diamond}\\
  & &  48 & 10 & - & Cree &  \cite{Wu_2004}\\
 \hline
 Pulsed & GaN & 15 & 2.5 & 25 & Lehigh & \cite{Luo_2011}\\
  & & 48 & 14.2 & 48 & CAS & \cite{Peng_2011}\\
  & & 28 & 6.43 & 42.5 & SEU Nanjing &  \cite{Tao_2016}\\
  & & 26 & 3.5 & 38 & MEC & \cite{Resca_2014}\\
  & & 20 & 3.68 & 59.9 & CU Boulder & \cite{Schafer_2013}\\
  & & 20 & 1.99 & 66.8 & &  \\
  & & 20 & 2.69 & 55.7 & &  \\
  & & 20 & 2.64 & 69.4 & &  \\
  & & 25 & 4 & 55 & Alcatel-Thales III-V Lab & \cite{Piotrowicz_2010}\\
  & & 25 & 7.1 & 51 & &  \\
  & & 32 & 6.5 & 38 & Alcatel-Thales III-V Lab & \cite{Piotrowicz_2008}\\
  & & 25 & 4.46 & 38 & &  \\
 \hline
 CW & GaN XHEMT & 25 & 4.5 & 53 & BAE Systems & \cite{GaN_XHEMT}\\
  & & 35 & 6.7 & 51 & & \\
  & & 40 & 7.5 & 48 & & \\
  & & 50 & 9.4 & 40 & & \\
 \hline
 CW & AlN & 15 & 2.58 & 25.6 & Cornell & \cite{Austin_FirstRF}\\
  & & 17 & 2.89 & 21.7 & &\\
  & & 20 & 3.3 & 14.8 & & \\
 \hline
 CW & AlGaN & 25 & 3.7 & 61 & HRL & \cite{HRL_2012} \\
  & & 30 & 5 & 58 & & \\
 \hline
 Pulsed & AlGaN & 30 & 7.1 & - & Fujitsu & \cite{Fujitsu_first} \\
  & & 40 & 8.9 & - & & \\
  & & 50 & 11 & - & & \\
  & & 60 & 13 & - & & \\
  & & 70 & 15.2 & 49.1 & & \\
  & & 30 & 9.4 & - & Fujitsu & \cite{Fujitsu_Record} \\
  & & 40 & 12.2 & - & & \\
  & & 50 & 14.6 & - & & \\
  & & 60 & 16.9 & - & & \\
  & & 70 & 18.7 & - & & \\
  & & 80 & 20.8 & - & & \\
  & & 90 & 22.3 & - & & \\
  & & 100 & 23.4 & - & & \\
  & & 110 & 24.4 & 46.8 &  &  \\
 \hline
\end{tabular}
\caption{The output power density and power-added efficiency of various GaN HEMTs in the X-band, as reported in the literature. The values listed here were used to generate Figure 5(b) and 5(c).}
\end{table}

\clearpage
\bibliographystyle{MSP}
\bibliography{references}

\end{document}